\documentclass[
preprint,
preprint,
amsmath,amssymb,
aps, physrev,
pra,
,superscriptaddress,nofootinbib,floatfix,superscriptaddress,amssymb
]{revtex4}
\usepackage{caption}
\usepackage{graphicx}
\usepackage{subcaption}%
\usepackage{dcolumn}
\usepackage{bm}
\usepackage{float}

\begin{document}
	
	\title{Entropy of Schwinger pair production in time-dependent Sauter pulse electric field}

	\author{Zhi-Hang Yao}
	
	\author{Hong-Hao Fan}
	\affiliation{Key Laboratory of Beam Technology of the Ministry of Education, and School of Physics and Astronomy, Beijing Normal University, Beijing 100875, China}
	
	\author{Lie-Juan Li}
	\affiliation{School of Mathematics and Physics, Lanzhou Jiaotong University, Lanzhou 730070, China}
	
	\author{Hai-Bo Sang}
	\affiliation{Key Laboratory of Beam Technology of the Ministry of Education, and School of Physics and Astronomy, Beijing Normal University, Beijing 100875, China}
	
	\author{Bai-Song Xie \footnote{Corresponding Author: bsxie@bnu.edu.cn}}
	\affiliation{Key Laboratory of Beam Technology of the Ministry of Education, and School of Physics and Astronomy, Beijing Normal University, Beijing 100875, China}
	\affiliation{Institute of Radiation Technology, Beijing Academy of Science and Technology, Beijing 100875, China}
	\date{\today}

\begin{abstract}
We investigate entropy of electron-positron pair production in time-dependent Sauter pulse electric field. Both cases of pair longitudinal momentum only and full momentum consideration are examined. We further examine three types of entropy, one is the usual entanglement entropy $S_{\text{E}}$, the other two extensions are thermal distribution entropy $S_{\text{Th}}$, and that with the chemical potential correction, $S_{\text{Th,CP}}$. For short pulse, $S_{\text{E}}$ is higher than $S_{\text{Th}}$ and vice versa for long pulse. The chemical potential causes the single-particle average thermal distribution entropy $\frac{S_{\text{Th,CP}}}{N}$ to exhibit non-monotonic behavior, similar to the single-particle average entanglement entropy $\frac{S_{\text{E}}}{N}$ in the short-pulse range.
In the full momentum case, we calculate the thermal distribution entropy $S_{\text{Th, U}}$ via introducing the Unruh temperature as the local effective temperature. We find that both $S_{\text{Th, U}}$ and $S_{\text{E}}$ saturate asymptotically to the constant while the former has a larger asymptotic value. The results presented in this study reveals that the different entropies have some delicate relationships among them.
\end{abstract}

\maketitle

\section{Introduction}\label{sec1}

Quantum entanglement, as one of the most prominent non-classical features of quantum mechanics, has been the crucial intersection of quantum foundational theory, high-energy physics, and quantum information science since the proposal of the Einstein-Podolsky-Rosen paradox \cite{36,37,38,39,26,40,41,42,43,44}.
The essence of quantum entanglement lies in the nonlocal correlations between the states of subsystems in a composite system, which can be experimentally verified through the violation of Bell's inequalities \cite{45,46,20,48}.

In strong field physics, the Schwinger effect describes the process of electron-positron pair (EPP) production from vacuum fluctuations in strong electric fields \cite{1,2,3,4,5,6,7}. A strong background electric field can destabilize the vacuum, and the created particle-antiparticle pairs inherently possess entanglement properties \cite{12,13}. Their entanglement strength is closely related to external field parameters. Through external field manipulation, the efficiency of entanglement generation and the purity of quantum states can be optimized \cite{49,51,27}. Within the relativistic quantum field theory framework, the Unruh effect reveals that an accelerated observer perceives the vacuum as a thermal bath. The created particles in this process are entangled and exhibit a nonlinear dependence on the magnitude of the acceleration. \cite{49,50,51}.

External field manipulation is one of the main directions in entanglement research.
In the time-dependent pulsed electric fields, the modulation of external field parameters can induce local peaks in entanglement \cite{27,48}. The entanglement entropy between particles and antiparticles is highly sensitive to external field parameters \cite{17} and is suppressed with an increase in the number of pulses \cite{18}.

However, the above studies largely focus only on the distribution of particles' longitudinal momentum. In the cases of large pulse widths or multi-photon absorption, the transverse momentum sharing occurs among particles. When calculating entanglement entropy, it is necessary to consider the effects of transverse momentum. 
On the other hand, the influence of chemical potential on particle production rates and entanglement entropy in a thermal vacuum background is also non-negligible, which is particularly important under finite temperature or non-equilibrium conditions.
This work extends the investigation to three-dimensional momentum space and compares the results with the one-dimensional case. It also considers the influence of chemical potential on entanglement characteristics.

In our investigation, we consider the following background field
\begin{equation}
	A_{1}(t) = E_{0}\tau\tanh\left(\frac{t}{\tau}\right), \quad E(t) = \frac{E_{0}}{\operatorname{csch}^{2}\left(\frac{t}{\tau}\right)},
	\label{eq4}
\end{equation}
where $E_0$ is the field strength and $\tau$ is the pulse duration. In the background field of the above form, the Dirac equation has an analytical solution.

In Sec.~\ref{sec2}, we present the results for  $S_{\text{E,L}}$, the entanglement entropy in the 1+1 dimensional simplified longitudinal momentum case, and for $S_{\text{E,F}}$, the entanglement entropy in the full momentum space. These results are obtained using the slowly varying electric field approximation.
Both $S_{\text{E,L}}$ and $S_{\text{E,F}}$ tend to asymptotic values over time, $S_{\text{E,L}}$ exhibits non-monotonic evolution, whereas $S_{\text{E,F}}$ increases monotonically. In Sec.~\ref{sec4}, we give the momentum distribution of EPP under the two cases. The momentum distribution exhibits a form similar to a thermal distribution. In Sec.~\ref{sec5}, we incorporate the chemical potential as a correction term, and introduce the effective temperature to present the thermal distribution entropy $S_{\text{Th}}$. We found that $S_{\text{Th}}$ is smaller than the $S_{\text{E}}$ in the short-pulse range and larger than $S_{\text{E}}$ in long pulse.
Additionally, the chemical potential correction term causes the single-particle average thermal distribution entropy,  $\frac{S_{\text{Th,CP}}}{N}$, to exhibit non-monotonic behavior in the short-pulse range.

\section{Entropies of EPP Production in Sauter Pulse electric field}

\subsection{Entanglement Entropy $S_{\text{E,L}}$ and $S_{\text{E,F}}$}\label{sec2}

In quantum electrodynamics, the production of EPP in a strong vacuum electric field is accompanied by entanglement. The degree of this quantum entanglement can be quantified by the entanglement entropy. Neglecting the transverse momentum of EPP under a linearly polarized external field, the equivalence between Gibbs entropy and von Neumann entanglement entropy can be demonstrated  in Ref.~\cite{21}. The equivalence can likewise be extended to the 1+3 dimensional case.

For the simplified longitudinal momentum case, the entanglement entropy $S_{\text{E,L}}(t)$ \cite{21} is given by
\begin{equation}
	S_{\text{E,L}}(t) = \int \frac{d k_{1}}{2\pi} \left[ \left(1 - \left|\beta_{k_{1},t}\right|^{2}\right) \log \left(1 - \left|\beta_{k_{1},t}\right|^{2}\right) + \left|\beta_{k_{1},t}\right|^{2} \log \left(\left|\beta_{k_{1},t}\right|^{2}\right) \right],
	\label{eq5}
\end{equation}
where $k_{1}$ is the one-dimensional parallel momentum, $\beta_{k_{1},t^{*}}$ is the Bogoliubov coefficient, which is from the Bogoliubov transformation relating the creation and annihilation operators with the momentum mode $k_{1}$ at the initial time $t = -\infty$ to those at time $t = t^{*}$ \cite{21,22,23}.

\begin{figure}[h]
	\centering
	
	\begin{subfigure}[b]{0.49\textwidth}
		\centering
		\includegraphics[width=\linewidth]{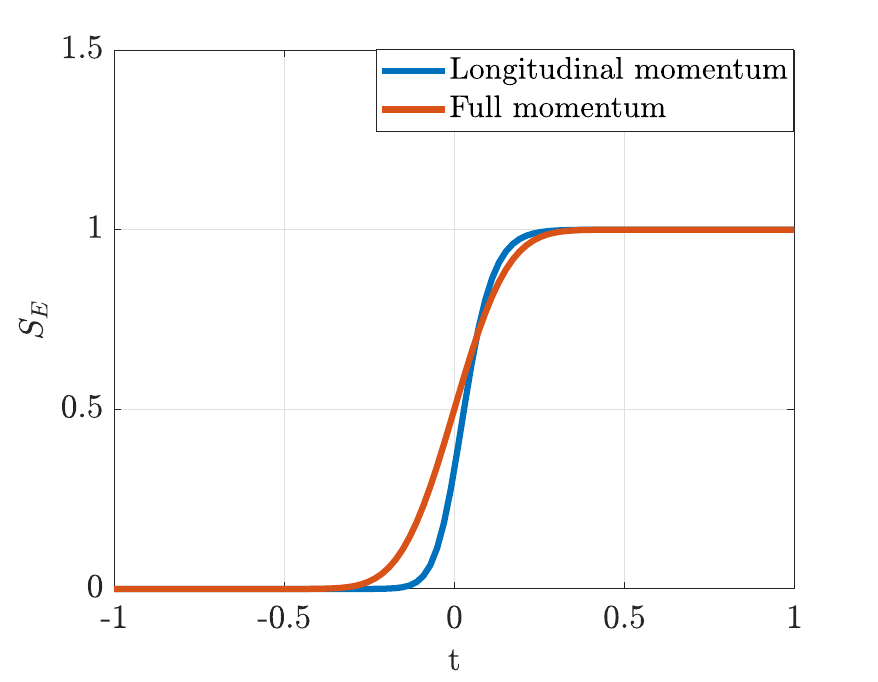}
		\caption{$\tau=1$}
		\label{subfig1a}
	\end{subfigure}
	\hspace{0pt}
	\begin{subfigure}[b]{0.49\textwidth}
		\centering
		\includegraphics[width=\linewidth]{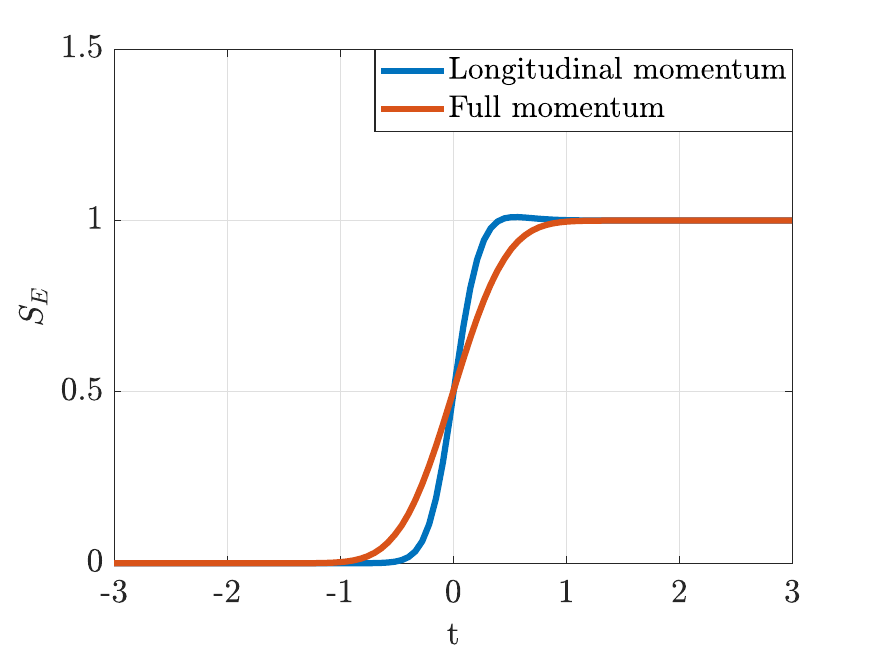}
		\caption{$\tau=3$}
		\label{subfig1b}
	\end{subfigure}
	
	\begin{subfigure}[b]{0.49\textwidth}
		\centering
		\includegraphics[width=\linewidth]{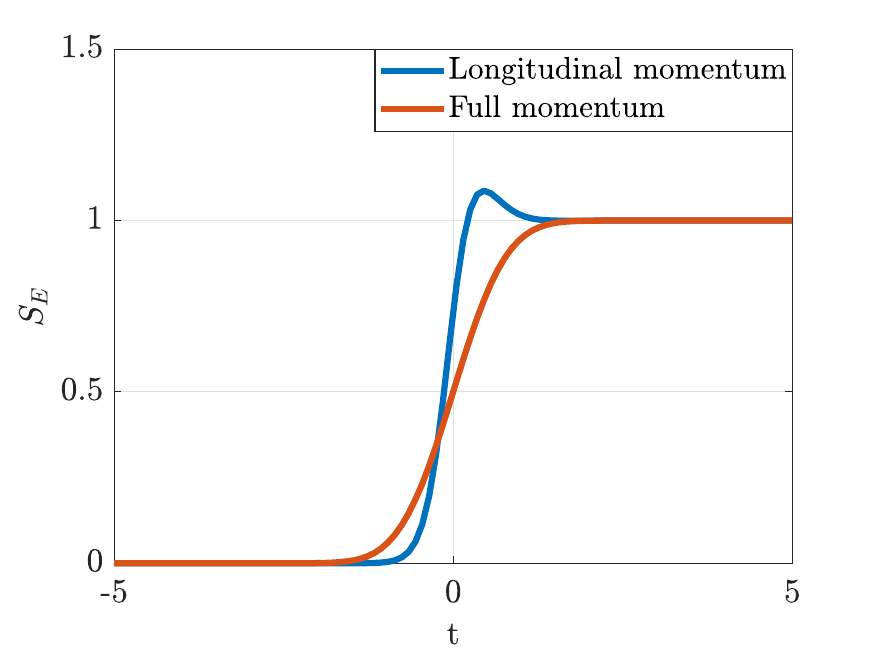}
		\caption{$\tau=5$}
		\label{subfig1c}
	\end{subfigure}
	\hspace{0pt}
	\begin{subfigure}[b]{0.49\textwidth}
		\centering
		\includegraphics[width=\linewidth]{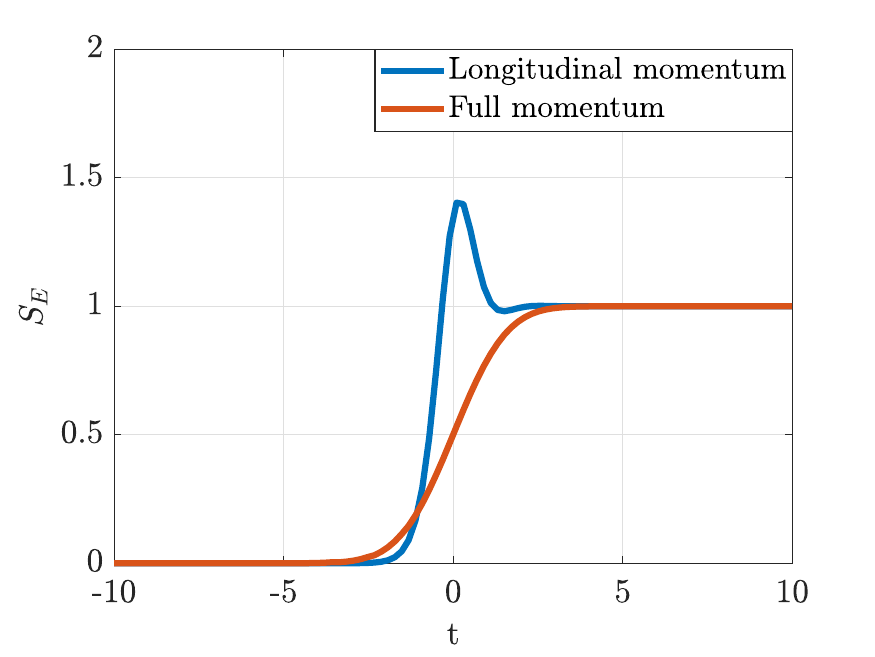}
		\caption{$\tau=10$}
		\label{subfig1d}
	\end{subfigure}
	\captionsetup{justification=raggedright}	
	\caption{Comparison of the time-dependent relationships of entanglement entropy between the longitudinal momentum case and the full momentum case.}
	\label{fig1}
\end{figure}

When calculating the expected number of electrons and positrons at time $t = t^{*}$, we find that the modulus squared of the Bogoliubov coefficient, denoted as $\left|\beta_{k_{1},t^{*}}\right|^{2}$, represents the probability of observing EPP in a momentum mode at time $t = t^{*}$. Therefore, $\left|\beta_{k_{1},t^{*}}\right|^{2}$ serves as the distribution function of EPP.
Using the Sauter pulse background field conditions and properties of the hypergeometric differential equation, we derive closed-form analytical solutions to the Dirac equation and the expression for $\left|\beta_{k_{1},t^{*}}\right|^{2}$ \cite{21,28}, which in turn makes $S_{\text{E,L}}(t)$ analytical. Using Eq.~\eqref{eq5}, we can derive the time dependence of $S_{\text{E,L}}(t)$, as shown by the blue curves in Fig.~\ref{fig1}.

To investigate the relevant behaviors of the entanglement entropy $S_{\text{E,L}}$, we introduce dimensionless adiabatic Keldysh parameter $\gamma = \frac{m}{e E_{0} \tau}$ \cite{29}. According to the conclusion in Ref.~\cite{21}, the entanglement entropy tends to an asymptotic value. As $\gamma$ increases, the entanglement entropy reaches equilibrium on a shorter time scale, with the peak-like feature in the time domain tends to vanish, see the blue curve in Fig.~ \ref{fig1}.

In the full momentum case, we focus on the case where the external field is a slowly varying Sauter pulse electric field. Therefore, we present the distribution function of EPP in the full momentum case under the condition of a slowly varying electric field as:
\begin{equation}
	f_{\text{Full}}(\boldsymbol{p},t)=\exp\left[-\pi\frac{E_{\text{c}}}{E(t)}\!\left(\!\frac{\varepsilon_\perp}{m}\!\right)^{\!\!2}\right],
	\label{eq10}
\end{equation}
where, $E_{\text{c}} = \frac{m^2}{e}$ is the critical electric field, and $\varepsilon_\perp = \sqrt{m^2 + \boldsymbol{p}_\perp^2}$ is the transverse energy.
Since the function in Eq.~\eqref{eq10} does not depend on the parameter $p_\parallel $, the integral over the entire space of $p_\parallel $ is not convergent and cannot be normalized. Therefore, we need to make a replacement: $\int dp_\parallel \to e\tau E_0$. Through this replacement, we can obtain the famous Schwinger formula \cite{3}. As the system evolves from the initial state to the final state, the entropy exhibits non-monotonic growth. Now the definition of entanglement entropy \cite{19} is given as:
\begin{equation}
	S_{\text{E,F}}(t)=-\int\frac{d \boldsymbol{p}}{(2 \pi)^{3}} f_{\text{Full}}(\boldsymbol{p}, t) \ln f_{\text{Full}}(\boldsymbol{p}, t)
	\label{eq11}.
\end{equation}
The above equation is equivalent to Eq.~\eqref{eq5}. This is because $f_{\text{Full}}(\boldsymbol{p},t) \ll 1$, so the term $(1-f_{\text{Full}}(\boldsymbol{p}, t)) \ln (1-f_{\text{Full}}(\boldsymbol{p}, t))$ can be neglected. Substituting Eq.~\eqref{eq10}, we can obtain the entropy production rate of the final state \cite{31} as:
\begin{equation}
	\frac{S_{\text{out}}}{T} = \frac{m^{4}}{8 \pi^{2}} \frac{E_{0}}{E_{c}} \left( 1 + \frac{E_{0}}{\pi E_{c}} \right) \exp \left( -\pi \frac{E_{c}}{E_{0}} \right).
	\label{eq12}
\end{equation}
Using the slowly varying electric field approximation, we can derive the time dependence of the entanglement entropy in the full momentum case with the slowly varying electric fields:
\begin{equation}
	S_{\text{E,F}}(t)=\int_{-\infty}^{t} dt\frac{m^{4}}{8\pi^{2}}\frac{E(t)}{E_{c}}\left(1+\frac{E(t)}{\pi E_{c}}\right)\exp\left(-\pi\frac{E_{c}}{E(t)}\right).
	\label{eq13}
\end{equation}
We take the same characteristic parameters ($E_0 = 0.1E_{\text{c}}$, $\tau = 1, 3, 5, 10$) for the longitudinal momentum case and the full momentum case. After normalization, we compare $S_{\text{E,L}}(t)$ and $S_{\text{E,F}}(t)$, as shown in Fig.~\ref{fig1}.

The longitudinal momentum and full momentum cases yield nearly overlapping curves, with the agreement between $S_{\text{E,L}}(t)$ and $S_{\text{E,F}}(t)$ strengthening as $\tau$ becomes smaller, and the peak feature of $S_{\text{E,L}}(t)$ tends to vanish. While $S_{\text{E,F}}(t)$ exhibits monotonic growth due to the slow-varying assumption of the time-varying electric field. Both $S_{\text{E,L}}(t)$ and $S_{\text{E,F}}(t)$ eventually converge to the stable values. The phenomenon can be primarily attributed to two reasons: First, the Pauli exclusion principle for fermions limits the particle occupation number of a single momentum mode, imposing an upper bound on the entanglement contribution of each momentum mode. Second, as the pulsed electric field is turned off, the production of EPP ceases, the particle number and momentum distribution no longer change, and entanglement loses its source of growth, thus stopping its rise and tending to a fixed value.

\subsection{The Momentum Dependence of EPP Production and the Thermal-like Behavior}\label{sec4}
In the simplified longitudinal momentum case, the asymptotic value of the entanglement entropy $S_{\text{E,L}}(\infty)$ can be calculated using the specific expression of the asymptotic coefficient \cite{21}. In order to compare the asymptotic state with the thermodynamic state, we expand the asymptotic coefficient for $\frac{1}{\gamma} \to 0$, and the asymptotic coefficient can be writed as \cite{21}:
\begin{equation}
	\lim _{\gamma \rightarrow \infty}\left|\beta_{k_{1}, \infty}\right|^{2} \sim
	\begin{cases}
		\left(\dfrac{E_{0} \tau}{m \pi}\right)^{2} \dfrac{1}{\left(m^{2}+k_{1}^{2}\right)^{2}}, & m \tau \ll 1 \\[2ex]
		\dfrac{4 m^{2} E_{0}^{2} \tau^{4}}{m^{2}+k_{1}^{2}} e^{-2 \pi \tau \sqrt{m^{2}+k_{1}^{2}}}, & m \tau \gtrsim 1
	\end{cases}.
	\label{eq15}
\end{equation}
Thus, we can obtain the distribution of the EPP production in the longitudinal momentum case over the parallel momentum $\Gamma_{\text{L}}(p_{\text{L}})=\left|\beta_{p_{\text{L}}, \infty}\right|^{2}$ \cite{21}. In the case of $m\tau \ll 1$, the momentum distribution decays in the form of $\frac{1}{(m^2 + k_1^2)^2}$. While in the case of $m\tau \gtrsim 1$, the momentum distribution exhibits a form similar to a thermal distribution \cite{24}, with $T_{\text{eff}} = \frac{1}{2\pi\tau}$ as the effective temperature, and there exists a slowly decaying factor in the form of $\frac{1}{m^2 + k_1^2}$.
Although we are currently dealing with the 1+1 dimensional case, the momentum distribution of EPP produced under short-pulse conditions is isotropic \cite{34}. In this case, the transverse momentum distribution is the same as the longitudinal momentum distribution.

Next, we consider the transverse momentum distribution of the EPP production in the full momentum case with slowly varying electric fields.
We can extract the dependence of created pair on transverse momentum by integrating the distribution function over the full range of parallel momentum
\begin{equation}
	\Gamma_{\text{F}}(\boldsymbol{p}_{\perp})= \int dt\, eE(t) \exp\left[ -\pi \frac{E_{\text{c}}}{E(t)} \left( \frac{\varepsilon_{\perp}}{m} \right)^2 \right].
	\label{eq16}
\end{equation}
Using Eq.~\eqref{eq16}, we can obtain the transverse momentum distribution of EPP in the full momentum case $\Gamma_{\text{F}}(\boldsymbol{p}_{\perp})$. We take the same characteristic parameters for $\Gamma_{\text{L}}(p_{\text{L}})$ and $\Gamma_{\text{F}}(\boldsymbol{p}_{\perp})$, normalize these two distributions respectively. $\Gamma_{\text{L}}(p_{\text{L}})$ and $\Gamma_{\text{F}}(\boldsymbol{p}_{\perp})$ are found to be completely identical, which indicates that when $\tau$ is large, $\Gamma_{\text{L}}(p_{\text{L}})$ and $\Gamma_{\text{F}}(\boldsymbol{p}_{\perp})$ exhibit the same thermal-like behavior. Thus, we can relate the asymptotic state to the thermodynamic state.

\subsection{Thermal Distribution Entropies $S_{\text{Th,B}}$, $S_{\text{Th,F-D}}$ and $S_{\text{Th,CP}}$}\label{sec5}

By using different thermal distributions, we can obtain the corresponding thermal distribution entropy $S_{\text{Th}}$ and compare it with $S_{\text{E}}$ under the Sauter pulse electric field.

In the longitudinal momentum case, we use the thermal distribution form of Eq.~\eqref{eq15} to obtain an effective temperature $T_{\text{eff}} = \frac{1}{2\pi \tau}$, and give the Boltzmann distribution $f_{\text{B}}(k_{1}) = e^{\frac{-\sqrt{m^2 + k_{1}^2}}{T_{\text{eff}}}}$. Through the effective temperature and the distribution function, we can calculate the thermal distribution entropy, and the steps are as follows:
\begin{equation}
	s = \frac{p+\epsilon}{T_{\text{eff}}},
	\label{eq17}
\end{equation}
with
\begin{align}
	\epsilon &= 2\int\frac{d k_{1}}{2\pi}\sqrt{m^{2}+{k_{1}}^{2}} f_{\text{B}}\left(k_{1}\right), \nonumber \\
	p &= 2\int\frac{d k_{1}}{2\pi}\frac{k_{1}}{\sqrt{m^{2}+k_{1}^{2}}} f_{\text{B}}\left(k_{1}\right), \nonumber \\
	n &= 2\int\frac{d k_{1}}{2\pi} f_{\text{B}}\left(k_{1}\right),
	\label{eq18}
\end{align}
where $s$, $p$, $\epsilon$, and $n$ denote the thermal distribution entropy, pressure, energy, and number density, respectively. $k_1$ is the one-dimensional parallel momentum.

\begin{figure}[htbp]
	\centering
	\includegraphics[width=0.9\textwidth]{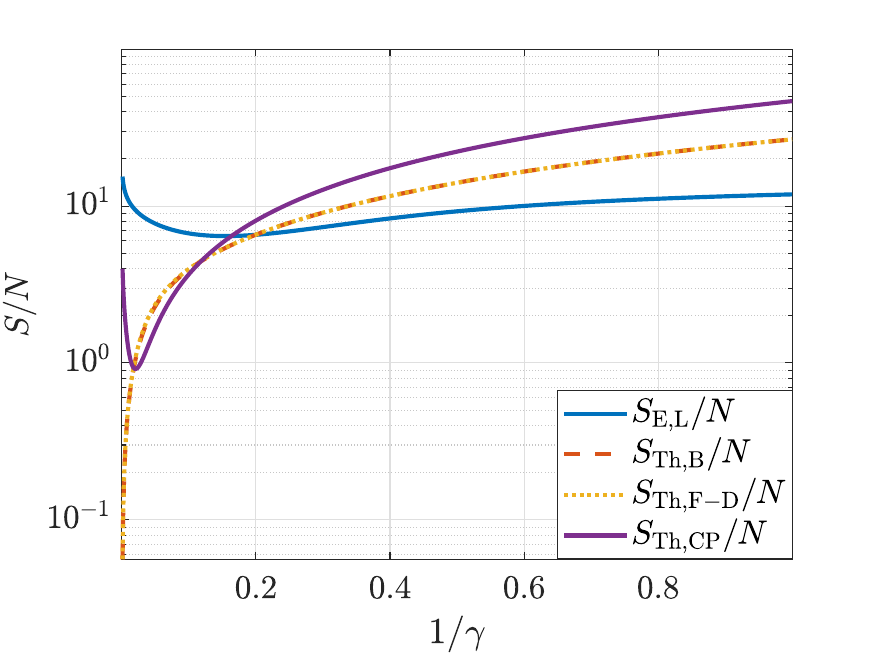}
	\captionsetup{justification=raggedright}	
	\caption{Comparison of entropis. The blue, red, yellow, and purple curves correspond to $\frac{S_{\text{E,L}}}{N}$, $\frac{S_{\text{Th,B}}}{N}$, $\frac{S_{\text{Th,F-D}}}{N}$, and $\frac{S_{\text{Th,CP}}}{N}$. We also reproduce the results of  Ref.~\cite{21} for the blue and red curves.}\label{fig4}
\end{figure}

Thus, we obtain the single-particle average Boltzmann entropy $\frac{S_{\text{Th,B}}}{N} = \frac{s}{n}$. We compare  $\frac{S_{\text{Th,B}}}{N}$ and the single-particle average entanglement entropy $\frac{S_{\text{E,L}}}{N}$ \cite{21} in Fig.~\ref{fig4}. The blue curve represents $\frac{S_{\text{E}}}{N}$ and the red curve represents $\frac{S_{\text{Th,B}}}{N}$; these two curves reproduce the results in Ref.~\cite{21}. On this basis, we further plot other curves in this paper.
We observe that over a specific range of pulse widths (near the intersection of the \(\frac{S_{\text{Th,B}}}{N}\) and \(\frac{S_{\text{E,L}}}{N}\) curves in Fig.~\ref{fig4}), \(\frac{S_{\text{E,L}}}{N}\) is nearly identical to \(\frac{S_{\text{Th,B}}}{N}\). In the limit of \( \gamma \to 0 \), \( \tau \to \infty \), i.e., the long-pulse case, $\frac{S_{\text{E,L}}}{N}$ tends to a constant value, while \( T_{\text{eff}} \to 0 \), \( p \to 0 \), \( \epsilon \sim mn \), \( \frac{s}{n} \sim \frac{m}{T_{\text{eff}}} \), $\frac{S_{\text{Th,B}}}{N}$ diverges, and $\frac{S_{\text{E,L}}}{N}$ is smaller than $\frac{S_{\text{Th,B}}}{N}$. When \( T_{\text{eff}} \to \infty \), $\frac{S_{\text{E,L}}}{N}$ is larger than $\frac{S_{\text{Th,B}}}{N}$. This situation arises because short pulses (\( \tau \to 0 \)) can only probe EPP in high frequency momentum modes; in this case, particle production originates from intense quantum fluctuations, and there exists strong entanglement between the created EPP. In the long-pulse range, $\frac{S_{\text{E,L}}}{N}$ and $\frac{S_{\text{Th,B}}}{N}$ behave similarly. However, an ideal thermal distribution represents a completely random state, while the created particles in the long-pulse electric field have quantum correlations and are not completely disordered and random, so $\frac{S_{\text{E,L}}}{N}$ is smaller than $\frac{S_{\text{Th,B}}}{N}$.

Compared with using the classical Boltzmann distribution, it may be more accurate to select the Fermi-Dirac distribution to statistically describe EPP. We replace the Boltzmann distribution \( f_{\text{B}}(k_1) \) with the Fermi-Dirac distribution $ f_{\text{F-D}}(k_1) = \left(1+ {\text{e}^{{\sqrt{m^2 + k_1^2}}/{T_{\text{eff}}}}} \right)^{-1}$, and calculate the corresponding single-particle average Fermi-Dirac entropy $\frac{S_{\text{Th,F-D}}}{N}$. In addition, we calculate the single-particle average Fermi-Dirac entropy with chemical potential $\frac{S_{\text{Th,CP}}}{N}$ by adding the chemical potential correction term as:
\begin{equation}
	s = \frac{p + \epsilon - \mu n}{T_{\text{eff}}},
	\label{eq19}
\end{equation}
with
\begin{equation}
	\mu = T_{\text{eff}} \log \left( \frac{n}{n_Q} \right), \quad	
	n_Q = \left( \frac{m T_{\text{eff}}}{2 \pi \hbar^{2}} \right)^{\frac{3}{2}},
	\label{eq20}
\end{equation}
where $\mu$ is the chemical potential .

We term the entropies derived from effective temperature and the distribution function as the thermal distribution entropy $S_{\text{Th}}$ for convenience. Based on the work presented in Ref.~\cite{21}, we additionally incorporate $\frac{S_{\text{Th,F-D}}}{N}$ and $\frac{S_{\text{Th,CP}}}{N}$ to yield the comprehensive Fig.~\ref{fig4}.
It can be observed that $\frac{S_{\text{Th}}}{N}$ remains smaller than $\frac{S_{\text{E,L}}}{N}$ in the short-pulse range, and remains larger than $\frac{S_{\text{E,L}}}{N}$ in the long-pulse range.

In Fig.~\ref{fig4}, $\frac{S_{\text{Th,F-D}}}{N}$ completely overlaps with $\frac{S_{\text{Th,B}}}{N}$. $\frac{S_{\text{Th,CP}}}{N}$ shows behavior similar to $\frac{S_{\text{E,L}}}{N}$ in the short-pulse range. In the short-pulse range, $\frac{S_{\text{Th,CP}}}{N}$ first decreases monotonically as the pulse width increases, and then begins to increase monotonically. In the long-pulse range, $\frac{S_{\text{Th,CP}}}{N}$ is larger than $\frac{S_{\text{Th,F-D}}}{N}$, indicating that the particle pairs in the low frequency momentum modes become more disordered when the chemical potential correction is taken into account.

In the full momentum case, since EPP carry electric charge, they will be accelerated in an external electric field, which leads to the Unruh effect \cite{35}, meaning that non-inertial observers in vacuum will perceive thermal radiation. In this case, we use the Unruh temperature $T_{\text{Unruh}}(t)$ as the effective temperature of the thermal distribution and compare it with the effective temperature $T_{\text{eff}}$ in the longitudinal momentum case, $T_{\text{Unruh}}(t) = \frac{E(t)}{E_{0} \gamma} T_{\text{eff}}$.
As we can see, $T_{\text{Unruh}}(t)$ is a local effective temperature that depends on the time-varying electric field. $T_{\text{eff}}$ is $T_{\text{Unruh}}(t)$ at the peak electric field multiplied by a coefficient $\gamma$. We use the $T_{\text{Unruh}}(t)$ to calculate the thermal distribution entropy of the full momentum case $S_{\text{Th,U}}(t)$.

\begin{figure}[h]
	\centering
	\includegraphics[width=0.9\textwidth]{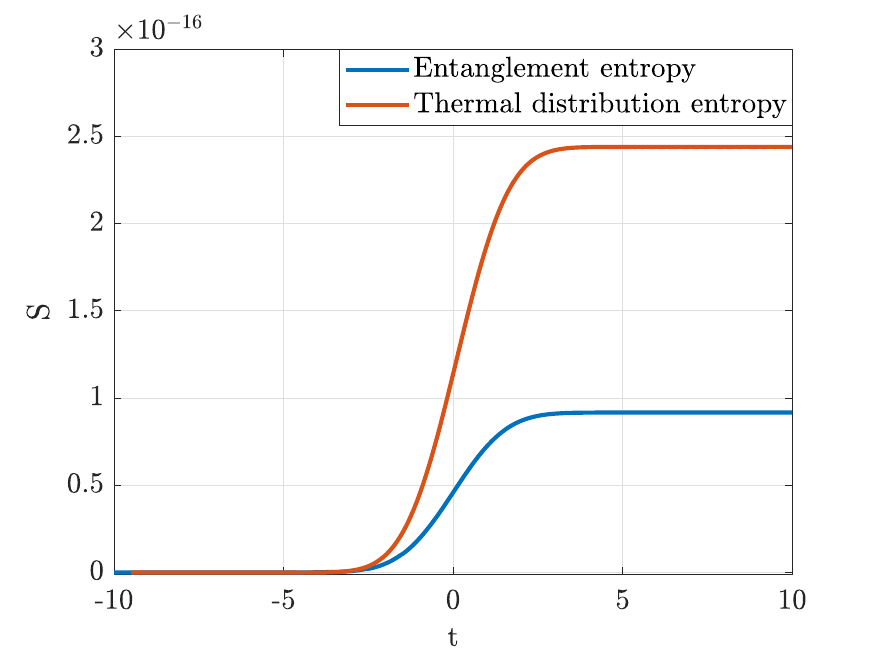}
	\caption{Comparison of $S_{\text{Th,U}}(t)$ and $S_{\text{E,F}}(t)$.}\label{fig5}
\end{figure}

Considering that the production of EPP is a non-Markovian process, where the current temperature still affects subsequent processes, we adopt a simple assumption: we use the time-dependent EPP production rate in the full momentum case $n_{\text{F}}(t)$ to perform a weighted average of $T_{\text{Unruh}}(t)$ at the current moment.
 We can obtain $n_{\text{F}}(t)$ by integrating $f_{\text{Full}}(\boldsymbol{p}, t)$ in momentum space
\begin{equation}
	n_{\text{F}}(t)  =\int_{-\infty}^{t} dt_{1}eE(t_{1})\int dp_{\perp} 2\pi p_{\perp}\exp\left[-\pi\frac{E_{\text{c}}}{E(t_{1})}\left(\frac{\varepsilon_{\perp}}{m}\right)^{2}\right].
\label{eq14}
\end{equation}
Thus, we obtain a non-Markovian average effective Unruh temperature $T_{\text{ave,eff}}(t)$ as follows:
\begin{equation}
	T_{\text{ave,eff}}(t) = \frac{1}{\int_{-\infty}^{t} n_{\text{F}}(t_{1}) dt_{1}} \int_{-\infty}^{t} dt_{1} n_{\text{F}}(t_{1}) T_{\text{Unruh}}(t_{1}).
	\label{eq22}
\end{equation}
We use $f_{\text{Full}}(\boldsymbol{p}, t)$ and $T_{\text{ave,eff}}(t)$ to calculate $S_{\text{Th,U}}(t)$. The calculation process is as follows:	
\begin{equation}
s = \frac{p + \epsilon - \mu n}{T_{\text{ave, eff}}(t)},	
\label{eq23}
\end{equation}
where $\epsilon$, $p$ and $n$ can be obtained when we make the replacements of $\frac{dk_1}{2\pi} \to \frac{d^3\boldsymbol{p}}{(2\pi)^3}$ and $f_\text{B}(k_1)\to f_{\text{Full}}(\boldsymbol{p}, t)$ in Eq.~\eqref{eq18}.

Through the above calculation, we can obtain the time-dependent relationship of the thermal distribution entropy for the full momentum $S_{\text{Th,U}}(t)$, as shown by the red curve in Fig.~\ref{fig5}, where the characteristic parameters are set as $E_0 = 0.1E_{\text{c}}$ and $\tau = 10$. It can be seen that when the pulsed electric field is turned on, $S_{\text{Th,U}}(t)$ starts from zero and continuously increases until it gradually reaches equilibrium near the field turn-off, eventually approaching an asymptotic value.

In Fig.~\ref{fig5}, we compare $S_{\text{Th,U}}(t)$ with $S_{\text{E,F}}(t)$. Since we consider the slowly
varying electric field, i.e., the long-pulse condition, the asymptotic value of $S_{\text{Th,U}}(t)$ is slightly larger than that of $S_{\text{E,F}}(t)$. A detailed explanation of this behavior is discussed in the analysis of Fig.~\ref{fig4}.

\section{Conclusion and Discussion}\label{sec6}

This study investigates the entropies during Schwinger pair production under a Sauter pulse electric field, systematically presents the calculation process of entanglement entropy and thermal distribution entropy, and reveals the profound connection between quantum entanglement and thermodynamic statistical behavior by comparing the longitudinal momentum case and the full momentum case. The main conclusions are as follows:

In the 1+1 dimensional case, the equivalence between Gibbs entropy and von Neumann entanglement entropy is strictly proven via Bogoliubov transformation, and this conclusion can be extended to the 1+3 dimensional case. The time evolution of $S_{\text{E}}$ shows that it tends to asymptotic equilibrium faster as the parameter $\gamma$ increases. Under short pulses ($\gamma \to \infty$), $S_{\text{E}}$ is higher than $S_{\text{Th}}$ due to the dominance of quantum fluctuations; under long pulses ($\gamma \to 0$), it is lower than $S_{\text{Th}}$ due to the limitation of quantum correlations. This reflects the difference between non-equilibrium dynamics and thermodynamic limits.

$S_{\text{E,L}}$ and $S_{\text{E,F}}$ are highly consistent under the slowly
varying electric field assumption, indicating that the simplified case can effectively capture the essential physical features of high-dimensional case. Momentum distribution analysis further confirms that the final-state EPP distribution exhibits thermal-like behavior; the quantitative calculation of $S_{\text{Th}}$ is achieved using the effective temperature $T_{\text{eff}}$ and the Unruh temperature $T_{\text{Unruh}}$.

By comparing with $\frac{S_{\text{Th,D-F}}}{N}$ and $\frac{S_{\text{Th,CP}}}{N}$, it is found that $\frac{S_{\text{E,L}}}{N}$ approximates the $\frac{S_{\text{Th}}}{N}$ under specific pulse parameters, but the chemical potential correction significantly affects the entropy behavior in the short-pulse range. 

If we want to accurately calculate entropy using a method similar to thermal distribution entropy, there are two key points: first, the selection of an appropriate effective temperature; second, a proper representation of the chemical potential. The expressions for the effective temperature and chemical potential in this paper are relatively rough, but they can already well fit the behavior of the entanglement entropy during the EPP production process. To obtaining a more accurate result, using a time- and momentum-dependent effective temperature $T_{\text{eff}}(t, \boldsymbol{p})$ in the calculation process could be a feasible approach.

This study provides a new perspective for entanglement quantification in strong-field quantum electrodynamics, emphasizes the key role of effective temperature and chemical potential in describing the statistical properties of non-equilibrium quantum systems.

\section*{Acknowledgements}

We are grateful to Q. Z. Lv for helpful discussions.
This research was supported by the National Natural Science Foundation of China (Grant Nos. 12375240, 12535015), and the Gansu Province Youth Science and Technology Fund (Grant No. 24JRRA276).

\section*{Declarations}
\begin{itemize}
\item Funding

National Natural Science Foundation of China Funded Project (12375240, 12535015)

Gansu Province Youth Science and Technology Fund (24JRRA276)
\end{itemize}

\bibliography{sn-bibliography}

\end{document}